# Three-dimensional solitons in Rydberg-Dressed cold atomic gases with spin-orbit coupling

Yuan Zhao [1,2,&], Heng-Jie Hu[1,3,&], Qian-Qian Zhou[1,3], Zhang-Cai Qiu[1,3], Li Xue[1,3], Si-Liu Xu [1,2*], Qin Zhou[4], and Boris A. Malomed[5,6]

[1]*Laboratory of Optoelectronic Information and Intelligent Control, Hubei University of Science and Technology, Xianning 437100, China*

[2] *School of Biomedical Engineering and Medical Imaging, Xianning Medical College, Hubei University of Science and Technology, Xianning 437100, China*

[3] *School of Electronic and Information Engineering, Hubei University of Science and Technology, Xianning 437100, China*

[4]*Research Center of Nonlinear Science, School of Mathematical and Physical Sciences, Wuhan Textile University, Wuhan 430200, China*

[5]*Department of Physical Electronics, School of Electrical Engineering, Faculty of Engineering, Tel Aviv Univesity, P.O.B. 39040, Ramat Aviv, Tel Aviv, Israel*

[6]*Instituto de Alta Investigación, Universidad de Tarapacá, Casilla 7D, Arica, Chile*

[&] *The authors contribute equally in this work*

*\* Corresponding author: xusiliu1968@163.com*

## Abstract

We present numerical results for three-dimensional (3D) solitons with symmetries of the semi-vortex (SV) and mixed-mode (MM) types, which can be created in spinor Bose-Einstein condensates of Rydberg atoms under the action of the spin-orbit coupling (SOC). By means of systematic numerical computations, we demonstrate that the interplay of SOC and long-range spherically symmetric Rydberg interactions stabilize the 3D solitons, improving their resistance to collapse. We find how the stability range depends on the strengths of the SOC and Rydberg interactions and the soft-core atomic radius.

## 1. Introduction

The formation of three-dimensional (3D) nonlinear localized states is a problem of fundamental importance in diverse areas of physics. It keeps drawing much interest, revealing various mechanisms which support the formation of such states [1-5]. The creation of 3D solitons is a significantly more challenging problem than the generation of low-dimensional ones, as the usual focusing cubic nonlinearity causes the critical and supercritical wave collapse in 2D and 3D geometries, respectively [6,7]. Various methods have been elaborated, chiefly in a theoretical form, to remedy this situation and stabilize multidimensional solitons, in the form of fundamental, multipolar, and vortical states [8,9]. Stable 3D solitons may be formed in settings with saturable or competing nonlinearities [10-12], nonlocal interactions [13,14], spatially modulated nonlinearity [15], *PT*-symmetric optical lattices [16,17], waveguide arrays and lattices imprinted in different materials [18-20], and binary Bose-Einstein condensates (BECs) subject to the action of spin-orbit coupling (SOC), which lend the system specific symmetry properties [21,22]. Accordingly, many kinds of solitons in SOC BECs, defined by their intrinsic symmetry, were predicted, such as semi-vortices (SVs) and mixed modes (MMs) [23-25], vortex-bright solitons [26-28], and filled-core vortices [29].

In particular, SOC BECs, composed as mixtures of atoms in two different hyperfine states, demonstrate coupling between the pseudospin degree of freedom and spatial structure of the condensate [5,30-32]. SOC notably modifies the dispersion of the system [33,34], breaks the Galilean invariance [35], and thus substantially impacts the properties of solitons in the free space [36,37]. Quite interesting is also the impact of SOC on BEC in external potentials, where possible symmetries of self-sustained solitons and their dynamics are determined by the symmetry of the potential [38-42]. A conclusion is that BECs under the action of SOC offer a versatile platform for the investigation of nonlinear phenomena in the presence of synthetic fields [43] and gauge potentials [44].

Theoretical [45] and experimental [46] studies have revealed that strong effective nonlinearities can be induced by the long-range Rydberg-Rydberg interaction (RRI) between remote atoms. To this end, RRI is mapped into a nonlocal optical nonlinearity through electromagnetically-induced transparency (EIT) at the single-photon level [47]. This option provides an important platform for the study of optical soliton dynamics with tunable parameters [48,49]. In particular, Rydberg gases are proven to be an effective medium for generating stable solitons with low energies under the action of the strong long-range RRI-induced nonlinearity [50-54]. Rydberg atomic gases are controllable in an active way through tunable parameters [55,56], such as atomic levels, detuning, laser intensities, etc. Furthermore, long lifetimes of the Rydberg atomic states (~ tens of microseconds) guarantee that the induced nonlinearities are quite robust [57]. Thus, Rydberg-EIT settings provide a fertile ground for realizing quantum nonlinear optics [48] and developing new photon devices, such as single-photon switches and transistors [58,59], quantum memories, and phase gates [56,60].

Although studies of various solitons in the context of BEC constitute a mature field, the existence and stability of the solitons in the framework of the mean-field theory, which is based on Gross-Pitaevskii equations (GPEs) under the action of SOC, contact interactions, and long-range spherically symmetric attractive RRIs remain an area of active work. The present paper aims to predict 3D solitons in binary atomic condensates combining SOC and RRI, taking into account the underlying symmetries and utilizing systematic numerical simulations of the corresponding model. In particular, we construct the solitons of the SV or MM types. Stability regions for the solitons are identified via the linear stability analysis and direct simulations of their perturbed evolution.

## 2. The Model and Numerical Method

As said above, the subject of the analysis is the 3D Rydberg-dressed binary BEC under the action of SOC. The scheme of the respective three-level atomic system is shown in Fig. 1(a). We consider the ultra-cold atomic gas of $N$ atoms, each possessing a ground state $|g\rangle$ and an excited one $|n'P\rangle$. An excited Rydberg state, composed of two Rydberg atoms at a distance $\mathbf{r}_{ij} = \mathbf{r}_i - \mathbf{r}_j$, is denoted $|e\rangle$. Two probe pulses with Rabi frequency $\Omega_{1/2}$ and detuning $\Delta_{1/2}$ couple the three-level states. In the mean-field approximation, the dynamics of the 3D spinor wave function $\Psi = (\Psi_+, \Psi_-)$ in the binary BEC with RRI and SOC of the Rashba type with strength $\lambda$ [61] is governed by the system of coupled scaled GPEs, which are written in the scaled form:

$$i\frac{\partial \Psi_+}{\partial t} = -\frac{1}{2}\nabla^2\Psi_+ + \left(|\Psi_+|^2 + |\Psi_-|^2\right)\Psi_+ + V_1\Psi_+ - i\lambda(\partial_x\Psi_- - i\partial_y\Psi_- + \partial_z\Psi_+) \quad (1a)$$

$$i\frac{\partial \Psi_-}{\partial t} = -\frac{1}{2}\nabla^2\Psi_- + \left(|\Psi_-|^2 + |\Psi_+|^2\right)\Psi_- + V_2\Psi_- - i\lambda(\partial_x\Psi_+ + i\partial_y\Psi_+ - \partial_z\Psi_-) \quad (1b)$$

where $\nabla^2 \equiv \partial_x^2 + \partial_y^2 + \partial_z^2$, the coefficients of the contact self- and cross-repulsion are set to be 1 by scaling (in most cases, these coefficients are nearly equal). The last term is the Rashba SOC which is obtained as $V_{\text{SOC}} = \lambda \mathbf{p} \cdot \boldsymbol{\sigma}$ [62], where $\mathbf{p} = -i\nabla$ is the momentum operator and $\boldsymbol{\sigma} = (\sigma_x, \sigma_y, \sigma_z)$ are Pauli matrices. The potential $V_1 = -\int d^3\mathbf{r}' \left[U_{11}(\mathbf{r}')|\Psi_+|^2 + U_{12}(\mathbf{r}')|\Psi_-|^2\right]$ and $V_2 = -\int d^3\mathbf{r}' \left[U_{21}(\mathbf{r}')|\Psi_+|^2 + U_{22}(\mathbf{r}')|\Psi_-|^2\right]$ represent the interaction between remote Rydberg atoms with

$$U_{ij} = C_{ij}/(R_c^6 + \mathbf{r}_{ij}^6) \quad (1c)$$

is the van der Waals potential, where $R_c$ is the soft-core radius, and

$$C_{ij} = (\Omega/2\Delta)^4 C_6^{(ij)}, \quad (1d)$$

where $C_6^{(ij)}$ are dispersion parameters which determine the intra- and inter-species couplings in the two-component BEC system [15,16]. Here $\Omega = \frac{\Omega_1\Omega_2}{2\Delta_1}$ is the effective Rabi frequency of the system, and $\Delta = \Delta_1 + \Delta_2$, where $\Omega_1$ and $\Omega_2$ are Rabi frequencies of two laser fields, $\Delta_1$ and $\Delta_2$ being the respective detunings. Note that, for a small Rydberg radius ($R_c \ll r_{ij}$), Eq. (1c) takes the form of $U_{ij} \approx C_{ij}/r_{ij}^6$, i.e., the RRI potential may be treated as a spherically symmetric *s*-wave scattering pseudopotential. In this work, by tuning system parameters, the values of $C_{ij}$, viz., $C_{11}$, $C_{12} = C_{21}$ and $C_{22}$, range from -1000 to 1000.

Our main focus herein is on discussing the effects of RRI and SOC on the formation of solitons. For $\Delta_1 \gg \Omega_1$, the system reduces to an effective two-level atom, with states $|g\rangle = |n_0 S\rangle$ and $|e\rangle = |nS\rangle$ coupled by a two-photon Rabi frequency $\Omega$ and detuning $\Delta$. For experimental considerations, a suitable candidate for the realization of the setup is the gas of $^{87}$Rb atoms with $\Delta = 2\pi \times 32 \text{MHz}$ and $\Omega = 2\pi \times 1 \text{MHz}$ [12,63].

Characteristic physical units related to the scaled ones in Eq. (1) are chosen as follows. The number of atoms in the binary BEC is characterized as $\approx 10^4 N$ where $N$ is the scaled norm of the wave function [64-66]. The scaled length and time units are $10\,\mu\text{m}$ and $100\,\text{ms}$, respectively. For example, $(x, y) = (5,5)$ corresponds to the spatial domain of size $(50\,\mu\text{m}, 50\,\mu\text{m})$, and $t = 10$ corresponds to $1000\,\text{ms}$. A typical distance between two Rydberg atoms is of the order $r_{ij} \approx 10\,\mu\text{m}$, while the unit for softcore size is estimated as $\approx 5.6\,\mu\text{m}$ [66].

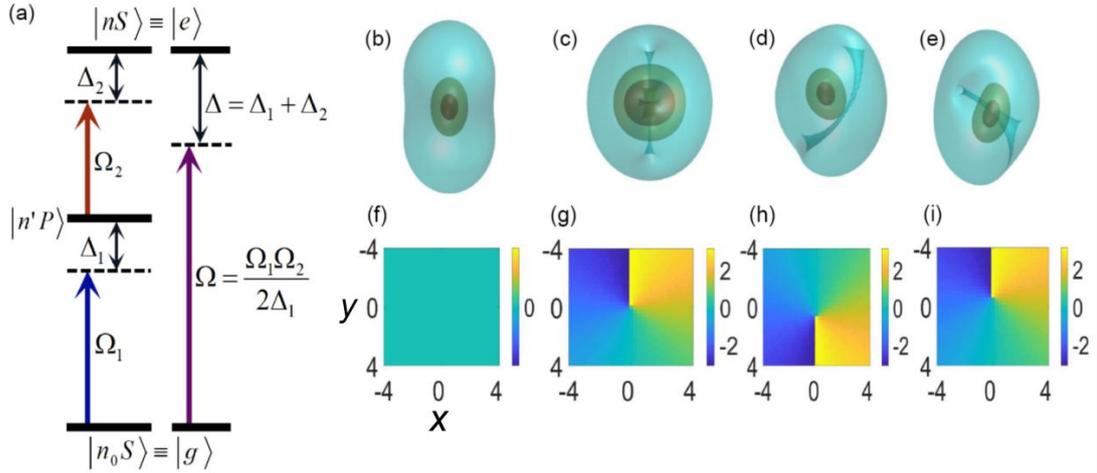

Fig. 1 (a) The schematic of the Rydberg-dressed three-level atomic system, with the laser coupling between the ground state $|n_0 S\rangle$ and the Rydberg state $|nS\rangle$. For $\Delta_1 \gg \Omega_1$, the system reduces to an effective two-level atom, with states $|g\rangle = |n_0 S\rangle$ and $|e\rangle = |nS\rangle$ coupled by the two-photon Rabi frequency $\Omega$ and detuning $\Delta$. Density isosurfaces of 3D solitons (b,c,d,e) and phase structures in $(x, y)$ plane (f,g,h,i) for the SV+, SV-, MM+, and MM- types.

Stationary spinor wave functions with chemical potential $\mu$ are looked for as $\Psi_\pm = \psi_\pm e^{-i\mu t}$. In cylindrical coordinates $(r, \varphi, z)$, solutions with symmetries corresponding to two different soliton species, SV and MM, are seeded by the following initial guesses for the stationary wave function [21]:

$$\psi_{SV+} = (A_1 + iB_1 z)e^{-\alpha_1 r^2 - \beta_1 z^2}, \quad \psi_{SV-} = (iA_2 + B_2 z)re^{-\alpha_2 r^2 - \beta_2 z^2 + i\varphi} \tag{2a}$$

for the SV (semi-vortex), and

$$\psi_{MM+} = \cos\theta \psi_{SV+} - \sin\theta \psi_{SV-}^*, \psi_{MM-} = \cos\theta \psi_{SV-} + \sin\theta \psi_{SV+}^* \tag{2b}$$

for the MM (mixed mode), where * stands for the complex conjugate.

The energy corresponding to Eqs. (1a) and (1b) is

$$\begin{aligned}
E_{tot} &= E_{kin} + E_{int} + E_{Ryd} + E_{SOC}, \\
E_{kin} &= \int \frac{1}{2}(|\nabla\psi_+|^2 + |\nabla\psi_-|^2)d\mathbf{r}, \\
E_{int} &= \int \left(\frac{1}{2}|\psi_+|^4 + \frac{1}{2}|\psi_-|^4 + |\psi_+|^2|\psi_-|^2\right)d\mathbf{r}, \\
E_{Ryd} &= -\iint d^3\mathbf{r}d^3\mathbf{r}'\left[\frac{1}{2}U_{11}(\mathbf{r}-\mathbf{r}')|\psi_+(\mathbf{r})|^2|\psi_+(\mathbf{r}')|^2 + \frac{1}{2}U_{22}(\mathbf{r}-\mathbf{r}')|\psi_-(\mathbf{r})|^2|\psi_-(\mathbf{r}')|^2 \right. \\
&\quad \left. + U_{12}(\mathbf{r}-\mathbf{r}')|\psi_+(\mathbf{r})|^2|\psi_-(\mathbf{r}')|^2\right], \\
E_{SOC} &= \lambda\int d^3\mathbf{r}\left(\psi_+^* \hat{D}_- \psi_- + \psi_-^* \hat{D}_+ \psi_+\right).
\end{aligned} \tag{3}$$

where SOC operators, with their respective symmetry structures, are $\hat{D}_\mp = \partial_x \mp i\partial_y \mp \partial_z$, and terms $E_{kin}$, $E_{int}$, $E_{Ryd}$, and $E_{SOC}$ represent the kinetic energy, inter- and intra-species interaction, RRI, and SOC interaction, respectively. Below, the RRI parameters $C_{ij}$, radius $R_c$ and SOC strength $\lambda$ are varied to study the soliton dynamics.

To explore the stability of solitons, perturbed solutions are taken as

$$\begin{aligned}
\Psi_+ &= (\psi_+ + g_1 e^{bt} + h_1^* e^{b^*t})e^{-i\mu t} \\
\Psi_- &= (\psi_- + g_2 e^{bt+ik\varphi} + h_2^* e^{b^*t-ik\varphi})e^{-i\mu t+i\varphi}
\end{aligned}, \tag{4}$$

where $g_{1,2}$ and $h_{1,2}$ are amplitudes of small perturbations, $b \equiv b_r + ib_i$ is the corresponding instability growth rate, and $k$ is the integer azimuthal index, $\varphi$ is the azimuthal angle. The ansatz for SVs permit both non-vortex and vortex modes, corresponding to the $\Psi_+$ and $\Psi_-$. For $\Psi_-$, the perturbation includes the azimuthal angle $\varphi$. The soliton solutions may be stable if $\text{Re}(b) = 0$ for all eigenvalues. The linearization of Eqs. (1) with respect to the small perturbations produces the system of

the respective Bogoliubov – de Gennes equations:

$$-ibg_1 = \left(\frac{1}{2}\nabla^2 + \mu - V_1 + i\lambda\frac{\partial}{\partial z} - |\psi_-|^2 - 2|\psi_+|^2\right)g_1 - \psi_+^2 h_1$$
$$+ \left[i\lambda\left(\frac{e^{ik\theta}\partial}{\partial x} + \frac{\partial e^{ik\theta}}{\partial x} - i\frac{e^{ik\theta}\partial}{\partial y} - i\frac{\partial e^{ik\theta}}{\partial y}\right) - \psi_+\psi_-^* e^{ik\theta}\right]g_2 - \psi_+\psi_- e^{ik\theta}h_2 \quad , \quad (5a)$$

$$-ibh_1 = \psi_+^{*2} g_1 + \left[-\frac{1}{2}\nabla^2 - \mu + V_1 + i\lambda\frac{\partial}{\partial z} + |\psi_-|^2 + 2|\psi_+|^2\right]h_1$$
$$+ \psi_+^*\psi_-^* e^{ik\theta} g_2 + \left[i\lambda\left(\frac{e^{ik\theta}\partial}{\partial x} + \frac{\partial e^{ik\theta}}{\partial x} + i\frac{e^{ik\theta}\partial}{\partial y} + i\frac{\partial e^{ik\theta}}{\partial y}\right) + \psi_+^*\psi_- e^{ik\theta}\right]h_2 \quad , \quad (5b)$$

$$-ibg_2 = \left[i\lambda\left(\frac{\partial}{\partial x} + \frac{e^{-ik\theta}\partial e^{ik\theta}}{\partial x} + i\frac{\partial}{\partial y} + i\frac{e^{-ik\theta}\partial e^{ik\theta}}{\partial y}\right) - \psi_+^*\psi_-\right]e^{-ik\theta}g_1 - \psi_+\psi_- e^{-ik\theta}h_1$$
$$+ \left(\frac{1}{2}\nabla^2 + \mu - V_2 - i\lambda\frac{\partial}{\partial z} - |\psi_+|^2 - 2|\psi_-|^2\right)g_2 - \psi_-^2 h_2 \quad , \quad (5c)$$

$$-ibh_2 = \psi_+^*\psi_-^* e^{-ik\theta} g_1 + \left[i\lambda\left(\frac{\partial}{\partial x} + \frac{e^{-ik\theta}\partial e^{ik\theta}}{\partial x} - i\frac{\partial}{\partial y} - i\frac{e^{-ik\theta}\partial e^{ik\theta}}{\partial y}\right) + \psi_+\psi_-^*\right]e^{-ik\theta}h_1$$
$$+ \psi_-^{*2} g_2 + \left(-\frac{1}{2}\nabla^2 - \mu + V_2 - i\lambda\frac{\partial}{\partial z} + |\psi_+|^2 + 2|\psi_-|^2\right)h_2 \quad . \quad (5d)$$

## 3. Results

In the present system, the total norm $N = \iiint (|\psi_+|^2 + |\psi_-|^2) dxdydz$ is composed of the spin-up and spin-down terms, $N^+ = \iiint |\psi_+|^2 dxdydz$ and $N^- = \iiint |\psi_-|^2 dxdydz$, respectively. They determine the respective norm shares, $F_\pm = N_\pm / N$.

Stationary solutions for the wave function seeded by the initial guesses (2) were obtained using the squared-operator method [67] and also by the imaginary-time evolution method[41]. As a result, two-component solitons are produced, SV={SV+,SV-}, and MM={MM+,MM-}. Examples of density isosurfaces and phase structure of these components of stable solitons are plotted in Figs. 1(b,c,d,e) and 1(f,g,h,i), respectively. For the mode with the SV symmetry, the norm is distributed along a ring, and the phase represents the vortex structure. On the other hand, in both components of the MM mode, the phase patterns are also vortical, while the density distributions seem like those representing "distorted vortices", cf.

qualitatively similar density patterns in the 3D SV and MM solitons produced by the GPE system with the contact-only cubic nonlinearity and SOC terms of the Weyl type[21].

**3.1 Families of the SV solitons**

Figure 2 summarizes the numerical solutions for the 3D solitons of the SV type, in the form of dependences of their energy, total norm, and the norm share of the SV+ component, on the Rydberg interaction strengths, $C_{ij}(i,j=1,2)$, and the softcore radius, $R_c$. It is seen that the energy, being vanishingly small for $C_{ij}<0$, rapidly increases with the increase of $C_{ij}>0$, seen in Fig. 2(a,b,c). Solid and dashed lines in Fig. 2 denote stable and unstable zones of the SV solutions in the panels. The results also indicate that the stability zone is smaller for larger $R_c$.

Stable SV solitons exist in an interval of the total norm

$$N_{\min} < N < N_{\max}, \tag{6}$$

which is shown in Figs. 2(d,e,f) for different values of $C_{ij}$. It is found that the dependences of $N_{\min}$ and $N_{\max}$ on the RRI coefficients $C_{11}$, $C_{12}$ and $C_{22}$ are completely different, demonstrating the absence of symmetry between different RRI constituents. The norm decreases monotonously with $C_{11}$, has a peak around $C_{12}=0$, and features a sharp drop at $C_{22}>250$.

As seen in Figs. 2(d,e,f), one finds that the total norm rapidly decreases at $C_{ij}>1$, indicating that the solitons in this BEC system vanish for large detunings. The norm share of the spin-up component of SVs is shown in Figs. 2(g,h,i). It increases monotonously with $C_{11}$ and is the dominant component of the norm at $C_{11}>500$. In contrast to the dependence on $C_{11}$, $F^+$ decreases monotonously with $C_{12}$ and $C_{22}$. Note that the increase of the Rydberg radius leads to a reduction of the stability zone of SVs, which is also observed in Fig. 2(a,b,c).

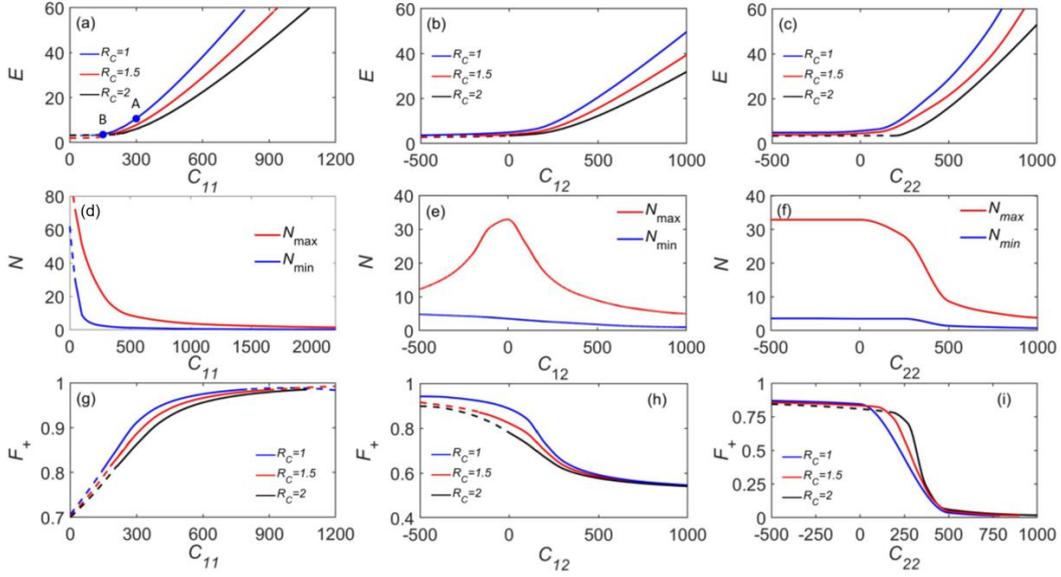

Fig. 2. The modulation of solitons of the SV type by altering the Rydberg coefficients, $C_{ij}$. The first to the third rows display, respectively, the total energy $E$ (a,b,c), the total norm $N$ of the solitons (d,e,f), and the norm share of the spin-up component, $F^+$ (g,h,i). The blue, red, and black lines pertain to Rydberg soft-core radii $R_c = 1.0, 1.5, 2.0$. Solid and dashed lines represent, respectively, quasi-stable and unstable solitons. Points A and B in panel (a) represent the stable and unstable states with $C_{11} = 300$ and 150, respectively. The fixed parameters are $\lambda = 1$, $C_{11} = 200$, $C_{12} = C_{21} = 0$, $C_{22} = 0$, $R_c = 1$ and $N = 5$.

Chemical potential $\mu$ as a function of $C_{ij}$ is plotted in Fig. 3(a,b,c) for different Rydberg soft-core radii $R_c$. It is seen that $\mu$ monotonously increases with $C_{11}$, and monotonously decreases with $C_{12}$ and $C_{22}$. The other difference between the dependences on $C_{11}$, $C_{12}$ and $C_{22}$ is the stability range. When $R_c = 1(5.6\mu m)$, the possibly stable solitons are relatively large, located in the ranges of $C_{11} \in (150,1000)$, $C_{12} \in (-500,1000)$ and $C_{22} \in (-500,1000)$. Note that, when the Rydberg soft-core radius is relatively large ($R_c = 2.0$), the stability zones for

$C_{12}$ and $C_{22}$ are much smaller.

The dependence of chemical potential $\mu$ on SOC strength $\lambda$ is shown in Fig. 3(g). It is seen that $\mu$ is much larger if the SOC is absent, $\lambda \approx 0$, but the solitons are unstable (dashed lines) in that case.

Similar to Fig. 2, solid and dashed lines in panels of Fig. 3 denote stable and unstable zones of the SV solutions. The stability zone of the solitons decreases dramatically with the growth of $R_c$. When $R_c = 1$, two stability zones are observed, divided by the area of $\lambda \approx 0$. With the increase of $R_c$, the stability zone shrinks. For $R_c = 2.0$ it is compressed into a very narrow area labeled by points A and B. This $R_c$ is a key factor that determines the soliton stability in the system. For the Rashba SOC, the actual particle current consists of both the canonical part, related to the superfluid velocity, and the SOC-induced gauge part, cf. Ref. [24]. Zhang *et al*. studied SOC BECs loaded into a toroidal trap and found that, for the counter-circling flow, these two parts have the same magnitude but opposite signs, creating a quasi-1D Rashba ring [68]. In the 3D BEC system with SOC, we find that the stable SV solitons and their chemical potentials show a nearly-symmetric response with respect to the substitution $\lambda \to -\lambda$. It is noticed that SV solitons cannot be generated at $|\lambda| < 0.6$.

The stability of the solitons may be also evaluated using the "anti-Vakhitov-Kolokolov" (anti-VK) criterion, $dN/d\mu > 0$, which is a necessary but not sufficient condition for the stability of solitons supported by repulsive (defocusing) nonlinearities [69]. To this end, the chemical potential $\mu$ is shown, as a function of $N$, in Figs. 3(d,e,f) for different $C_{ij}$, and in Fig. 3(h) for different $\lambda$, demonstrating that an anti-VK criterion holds. In the full form, the stability of SVs is determined by the eigenvalues for small perturbations, produced by the numerical solution of Eqs. (5). The real part of the eigenvalues is shown in Fig. 3(i). One observes that $\mathrm{Re}(b)$ is close to zero($\sim 10^{-5}$) in broad intervals, where the SVs are quasi-stable states. The stability of solitons of the MMs type is similarly determined by the eigenvalues.

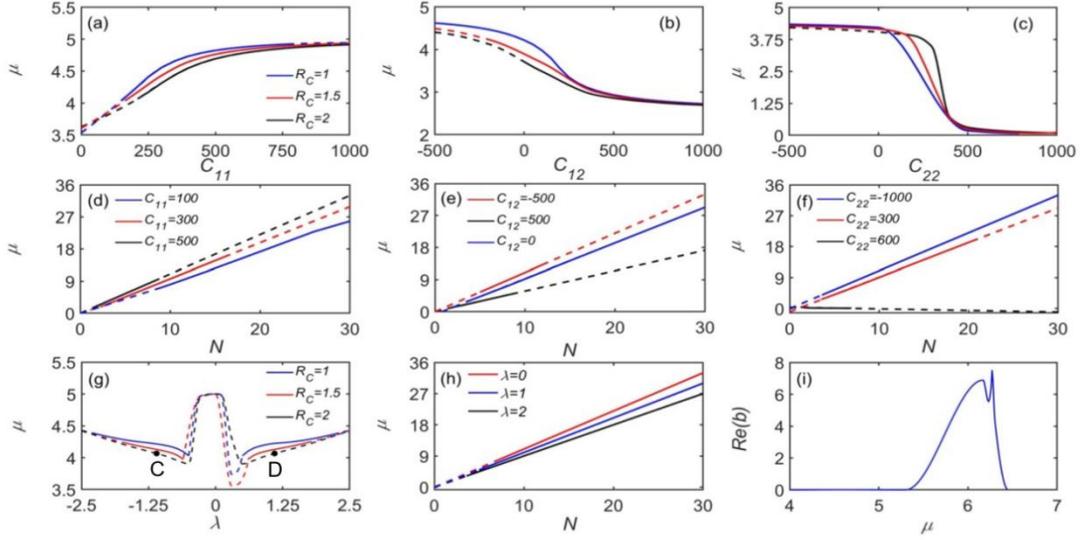

Fig. 3. Chemical potential $\mu$ for SVs with different values of the system's parameters, including $C_{ij}$, $N$, and $\lambda$. Solid and dashed lines represent, respectively, quasi-stable and unstable soliton families. (a,b,c) The first row is the $\mu - C_{ij}$ relation for different Rydberg soft-core radii $R_c = 1.0, 1.5, 2.0$. (d,e,f) The second row is the $\mu - N$ relation for different $C_{ij}$ when $R_c = 1.0$. (g) The $\mu - \lambda$ relation for different values of $R_c$, where points C and D stand for very narrow zones of quasi-stability at $R_c = 2.0$. (h) The $\mu - N$ relation for different $R_c$. Solid and dashed lines represent, respectively, stable and unstable solitons. (i) The real part of the stability eigenvalue vs. the chemical potential when $\lambda = 1$, $C_{11} = 200$, $C_{12} = C_{21} = 0$, $C_{22} = 0$, $R_c = 1$ and $N = 5$. Other parameters are the same as in Fig. 2.

To explore the dynamics of SVs, profiles of SVs and MMs were produced by direct simulations of their perturbed evolution. In Fig. 4, SV+ and SV- with two sets of parameters are shown. The magnitudes of the three-layer isosurface displayed in the figure are $(0.95, 0.5, 0.05)|\psi|_{\max}$, where $|\psi|_{\max}$ is the amplitude of the wave field. It is observed in Fig. 4 that the times during which the SVs maintain their integrity are different, depending on the initial inputs. In the first and second columns of Fig. 4, the soliton is quasi-stable, while in the third and fourth columns, it is definitely unstable.

The stability of these solitons was evaluated by the real part of the eigenvalue $\mathrm{Re}(b)$, as shown in Fig. 3(i). The SVs are generated at $t=0$ and evolve in time. $\mathrm{Re}(b)$ is different depending on the initial inputs. In Fig. 4, SV+ and SV- in the first and second columns demonstrate a small $\mathrm{Re}(b) = 2.0\times 10^{-5}$. However, it is larger in the third and fourth columns, $viz.$, $\mathrm{Re}(b) = 1.39\times 10^{-2}$. Accordingly, the respective soliton collapses faster. Thus, to maintain the soliton's stability (or keep the instability weak enough), it is important to optimize parameters of the setting.

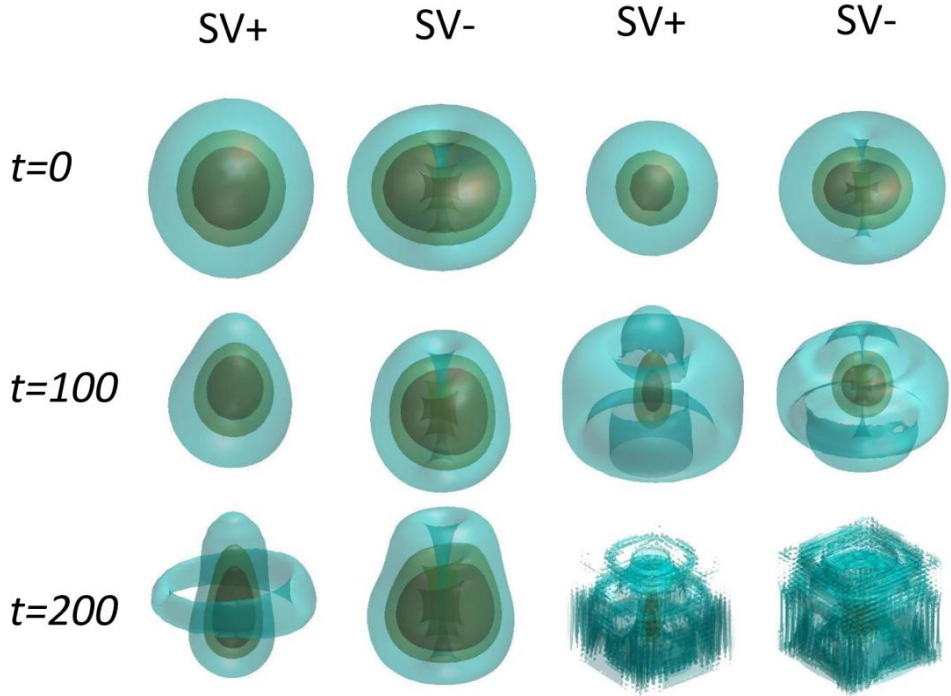

Fig. 4. Isosurfaces of the SV+ and SV- components as produced by the direct simulations. The first and second columns represent the quasi-stable SVs (point A in Fig. 2(a)) with $\mathrm{Re}(b) = 2.0\times 10^{-5}$, while the third and fourth columns pertain to strongly unstable SVs (point B in Fig. 2(a)) with $\mathrm{Re}(b) = 1.39\times 10^{-2}$.

## 3.2 Solitons of MM type

The dynamics of MMs is nearly the same as that of SVs. Figure 5 shows the perturbed evolution of the solitons of this type, shown by three-layer isosurface configurations. The (in)stability of these solitons is also characterized by $\text{Re}(b)$. The solitons displayed in the first (MM+) and second (MM-) columns of Fig. 5 are identified as quasi-stable ones with $\text{Re}(b) = 2.53 \times 10^{-5}$, while the third (MM+) and fourth (MM-) columns show their strongly unstable counterpart with $\text{Re}(b) = 5.85 \times 10^{-2}$. Thus, the quasi-stable and strongly unstable states are produced by the analysis.

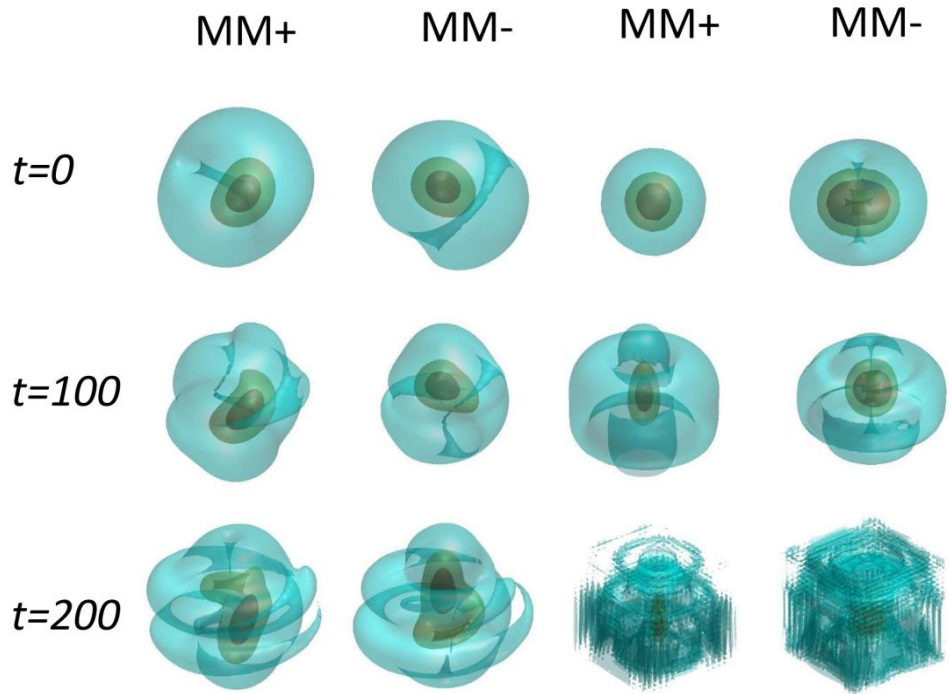

Fig. 5. Isosurfaces of the MM+ and MM- components, as produced by the direct simulations. The first and second columns represent the quasi-stable MMs for parameters $\lambda = 1$, $C_{11} = 300$, $C_{12} = C_{21} = 0$, $C_{22} = 0$, $R_c = 1$, the respective instability growth rate being $\text{Re}(b) = 2.53 \times 10^{-5}$, while the third and fourth columns pertain to $\lambda = 1$, $C_{11} = 100$, $C_{12} = C_{21} = 0$, $C_{22} = 0$, $R_c = 1$, and $\text{Re}(b) = 5.85 \times 10^{-2}$.

To quantify the evolution of the solitons, the average width in three directions is defined as

$$W_j(t) = \left( \frac{1}{N} \iiint j^2 \left[ |\psi_+(x,y,z;t)|^2 + |\psi_-(x,y,z;t)|^2 \right] dxdydz \right)^{1/2}, j = x, y, z. \quad (7)$$

Thus, the anisotropy of the solitons is defined as the width ratio $W_x/W_y$, and its area is $S = \pi W_x \times W_y$. These characteristics of the SVs and MMs are shown, as a function of time, in Fig. 6. It is seen that the width ratio keeps values $W_x/W_y < 1$ and $W_x/W_y > 1$ for the SV and MM modes, respectively. The area in the $(x, y)$ plane increases monotonously, indicating gradual spread of the wave functions in the course of the propagation. Though the SVs and MMs are not completely stable, their instability growth rates may be small, allowing long survival times. Such unstable nonlinear states can be regarded as practically stable objects, taking into regard time limitations in experiments.

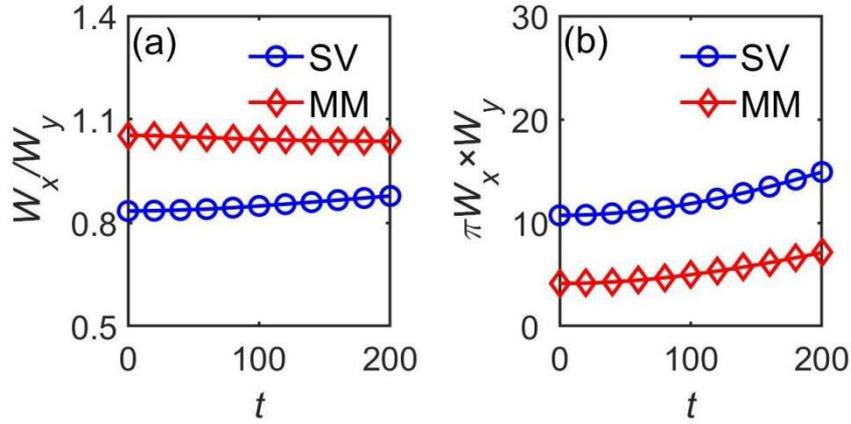

Fig. 6 The evolution of the asymmetry ratios and areas of the SV and MM solitons, defined as per Eq. (7), in the course of their propagation. (a) and (b) are width ratio $W_x/W_y$ and area $\pi W_x W_y$, respectively. The parameters for stable SVs and MMs are $\lambda = 1$, $C_{11} = 200$, $C_{12} = C_{21} = 0$, $C_{22} = 0$, $R_c = 1$.

## 4. Conclusion

The Rydberg-dressed binary BEC with SOC (spin-orbit coupling) is proposed here to produce quasi-stable 3D solitons. The three-level atomic scheme is constructed by coupling the two-level atomic structure to the excited Rydberg state. Three-dimensional Gross-Pitaevskii equations are introduced to govern the dynamics of the system. Solitons with SV (semi-vortex) and MM (mixed-mode) symmetries are obtained by tuning the system's parameters, such as Rydberg interaction coefficients $C_{ij}$, soft-core radius $R_C$, and SOC strength $\lambda$. These solitons are proven to be quasi-stable by means of the linearized analysis and direct simulations. The quasi-stability zones for the solitons of the SV and MM types are mainly determined by $R_C$, and the dynamics can be effectively controlled by $C_{ij}$ and $\lambda$.

The work can be extended for the consideration of interactions between solitons. It may also be relevant to analyze the possibility of the existence of light bullets in the SOC-Rydberg medium.

**Acknowledgment**

We thank Prof. P. G. Kevrekidis for valuable discussions.

**Funding**

National Natural Science Foundation of China (62275075, 11975172 and 12261131495); Natural Science Foundation of Hubei province (2023AFC042). Israel Science Foundation (1695/22).

**Author Contribution**

Y. Zhao wrote the main manuscript text. H.-J. Hu, Q.-Q. Zhou and Z.-C. Qiu done the calculation and prepared the figures. L. Xue, Q. Zhou and B.A. Malomed participated in the discussion and revision. S.-L. Xu provided the method and formula analysis. All authors reviewed the manuscript and contributed to the interpretation of the work.

**Data availability**

The datasets generated during and/or analysed during the current study are available from the corresponding author on reasonable request.

**Additional Information**

Competing financial interests: The authors declare no competing financial interests.

**References:**

[1] Malomed, B. A., Mihalache, D., Wise, F. & Torner, L. Spatiotemporal optical solitons. *J. Opt. B* **7**, R53-R72 (2005).

[2] Ackerman, P. J., & Smalyukh, I. I. Diversity of knot solitons in liquid crystals manifested by linking of preimages in torons and hopfions. *Phys. Rev. X* **7**, 011006 (2017).

[3] Kartashov, Y. V., Astrakharchik, G. E., Malomed, B. A., & Torner, L. Frontiers in multidimensional self-trapping of nonlinear fields and matter. *Nature Reviews Physics* **1**, 185-197 (2019).

[4] Mihalache, D. Multidimensional localized structures in optical and matter-wavemedia: A topical survey of recent literature, *Rom. Rep. Phys.* **69**, 403 (2017).

[5] Malomed, B. A. Multidimensional Solitons, American Institute of Physics Publishing, (Melville, NY, 2022).

[6] Bergé, L. Wave collapse in physics: principles and applications to light and plasma waves. *Physics Reports* **303**, 259-370(1998).

[7] Kuznetsov, E. A. & Dias, F.. Bifurcations of solitons and their stability. *Physics Reports* **507**, 43–105(2011).

[8] Mihalache, D. Formation and stability of light bullets: Recent theoretical studies. *Optoelectronics & Advanced Materials* **12**, 12-18 (2010).

[9] Chen, S.-F. *et al.* Vortex solitons in Bose-Einstein condensates with spin-orbit coupling and Gaussian optical lattices. *Appl. Math. Lett.* **92**, 15-21 (2019).

[10] Desyatnikov, A., Maimistov, A. & Malomed, B. Three-dimensional spinning solitons in dispersive media with the cubic-quintic nonlinearity. *Phys. Rev. E* **61**, 3107–3113 (2000).

[11] Mihalache, D. *et al.* Stable spinning optical solitons in three dimensions. *Phys.*

*Rev. Lett.* **88**, 073902 (2002).

[12] Falcão-Filho, E. L., de Araújo, C. B., Boudebs, G., Leblond, H. & Skarka, V. Robust Two-Dimensional Spatial Solitons in Liquid Carbon Disulfide. *Phys. Rev. Lett.* 110, 013901 (2013).

[13] Maucher, F., Henkel, N., Saffman, M., Królikowski, W., Skupin, S., & Pohl, T. Rydberg-Induced Solitons: Three-Dimensional Self-Trapping of Matter Waves. *Phys. Rev. Lett.* **106**, 170401 (2011).

[14] Tikhonenkov, I., Malomed, B. A., & Vardi, A. Anisotropic solitons in dipolar Bose-Einstein Condensates. *Phys. Rev. Lett.* **100**, 090406 (2008).

[15] Hsueh, C.-H., Tsai, Y.-C., Wu, K.-S., Chang, M.-S., & Wu, W. C. Pseudospin orders in the supersolid phases in binary Rydbergdressed Bose-Einstein condensates. *Phys. Rev. A* **88**, 043646 (2013).

[16] Han, W., Zhang, X.-F., Wang, D.-S., Jiang, H.-F., Zhang, W., & Zhang, S.-G. Chiral Supersolid in Spin-Orbit-Coupled Bose Gases with Soft-Core Long-Range Interactions. *Phys. Rev. Lett.* **121**, 030404 (2018).

[17] Li, J., Zhang, Y., & Zeng, J. Matter-wave gap solitons and vortices in three-dimensional parity-time-symmetric optical lattices. *iScience* **25**, 104026 (2022).

[18] Baizakov, B. B., Malomed, B. A., & Salerno, M. Multidimensional solitons in a low-dimensional periodic potential. *Phys. Rev. A* **70**, 053613 (2004).

[19] Milián, C., Kartashov, Y. V., & Torner, L. Robust Ultrashort Light Bullets in Strongly Twisted Waveguide Arrays. *Phys. Rev. Lett.* **123**, 133902 (2019).

[20] Mihalache, D. *et al.* Stable Spatiotemporal Solitons in Bessel Optical Lattices. *Phys. Rev. Lett.* **95**, 023902 (2005).

[21] Zhang, Y.-C., Zhou, Z.-W., Malomed, B. A., & Pu, H. Stable Solitons in Three Dimensional Free Space without the Ground State: Self-Trapped Bose-Einstein Condensates with Spin-Orbit Coupling. *Phys. Rev. Lett.* **115**, 253902 (2015).

[22] Kartashov, Y. V. & Zezyulin, D. A. Stable Multiring and Rotating Solitons in Two-Dimensional Spin-Orbit-Coupled Bose-Einstein Condensates with a Radially Periodic Potential. *Phys. Rev. Lett.* **122**, 123201 (2019).

[23] Sakaguchi, H., Li, B., & Malomed, B. A. Creation of two-dimensional composite

solitons in spin-orbit-coupled self-attractive Bose-Einstein condensates in free space. *Phys. Rev. E* **89**, 032920 (2014).

[24] Sakaguchi, H., Sherman, E. Y., & Malomed, B. A. Vortex solitons in two-dimensional spin-orbit coupled Bose-Einstein condensates: Effects of the Rashba-Dresselhaus coupling and Zeeman splitting. *Phys. Rev. E* **94**, 032202 (2016).

[25] Liao, B. *et al.* Anisotropic solitary semivortices in dipolar spinor condensates controlled by the two-dimensional anisotropic spin-orbit coupling, *Chaos, Solitons and Fractals* **116**, 424-432 (2018).

[26] Pola, M., Stockhofe, J., Schmelcher, P., & Kevrekidis, P. G. Vortex‐bright-soliton dipoles: Bifurcations, symmetry breaking, and soliton tunneling in a vortex-induced double well. *Phys. Rev. A* **86**, 053601 (2012).

[27] Gautam, S. & Adhikari, S. K. Vortex-bright solitons in a spin-orbit-coupled spin-1 condensate. *Phys. Rev. A* **95**, 013608 (2017)

[28] Law, K. J. H., Kevrekidis, P. G., & Tuckerman, L. S. Stable vortex-bright-soliton structures in two-component Bose-Einstein condensates. *Phys. Rev. Lett.* **105**, 160405 (2010).

[29] Anderson, B. P., Haljan, P. C., Wieman, C. E., & Cornell, E. A. Vortex Precession in Bose-Einstein Condensates: Observations with Filled and Empty Cores. *Phys. Rev. Lett.* **85**, 2857-2860 (2000)

[30] Lin, Y.-J., Jiménez-García, K. & Spielman, I. B. Spin-orbit-coupled Bose-Einstein condensates. *Nature (London)* **471**, 83-88 (2011).

[31] Hamner, C., Zhang, Y., Khamehchi, M. A., Davis, M. J., & Engels, P. Spin-Orbit-Coupled Bose-Einstein Condensates in a One-Dimensional Optical Lattice. *Phys. Rev. Lett.* **114**, 070401 (2015).

[32] Wu, Z. *et al.* Realization of two-dimensional spin-orbit coupling for Bose-Einstein condensates. *Science*, **354**, 83-88 (2016).

[33] Li, Y., Martone, G. I., Pitaevskii, L. P. & Stringari, S. Superstripes and the Excitation Spectrum of a Spin-Orbit-Coupled Bose-Einstein Condensate. *Phys. Rev. Lett.* **110**, 235302 (2013).

[34] Zhang, Y., Mao, L., & Zhang, C. Mean-Field Dynamics of Spin-Orbit-Coupled

Bose-Einstein Condensates. *Phys. Rev. Lett.* **108**, 035302 (2012).

[35] Zhu, Q., Zhang, C., & Wu, B. Exotic superfluidity in spin-orbit coupled Bose-Einstein condensates. *Europhys. Lett.* **100**, 50003 (2012).

[36] Achilleos, V., Frantzeskakis, D. J., Kevrekidis, P. G., & Pelinovsky, D. E. Matter-Wave Bright Solitons in Spin-Orbit Coupled Bose-Einstein Condensates. *Phys. Rev. Lett.* **110**, 264101 (2013).

[37] Xu, Y., Zhang, Y., & Wu, B. Bright solitons in spin-orbit-coupled Bose-Einstein condensates. *Phys. Rev. A* **87**, 013614 (2013).

[38] Fialko, O., Brand, J., & Zülicke, U. Soliton magnetization dynamics in spin-orbit-coupled Bose-Einstein condensates. *Phys. Rev. A* **85**, 051605(R) (2012).

[39] Zezyulin, D. A., Driben, R., Konotop, V. V., & Malomed, B. A. Nonlinear modes in binary bosonic condensates with pseudo spin-orbital coupling. *Phys. Rev. A* **88**, 013607 (2013).

[40] White, A. C., Zhang, Y., & Busch, T. Odd-petal-number states and persistent flows in spin-orbit-coupled Bose-Einstein condensates. *Phys. Rev. A* **95**, 041604(R) (2017).

[41] Li, H., Xu, S.-L., Belić, M. R., & Cheng, J.-X. Three-dimensional solitons in Bose-Einstein condensates with spin-orbit coupling and Bessel optical lattices. *Phys. Rev. A* **98**, 033827 (2018).

[42] Lobanov, V. E., Kartashov, Y. V., & Konotop, V. V. Fundamental, Multi-pole, and Half-Vortex Gap Solitons in Spin-Orbit Coupled Bose-Einstein Condensates. *Phys. Rev. Lett.* **112**, 180403 (2014).

[43] Lin, Y.-J., Compton, R. L., Jiménez-García, K., Porto, J. V., & Spielman, I. B. Synthetic magnetic fields for ultracold neutral atoms. *Nature (London)* **462**, 628-632 (2009).

[44] Ho, T.-L., & Zhang, S. Bose-Einstein Condensates with Spin-Orbit Interaction. *Phys. Rev. Lett.* **107**, 150403 (2011).

[45] Friedler, I., Petrosyan, D., Fleischhauer, M., & Kurizki, G. Long-range interactions and entanglement of slow single-photon pulses. *Phys. Rev. A* **72**, 043803 (2005).

[46] Mohapatra, A. K., Jackson, T. R., & Adams, C. S. Coherent optical detection of highly excited Rydberg states using electromagnetically induced transparency. *Phys. Rev. Lett.* **98**, 113003 (2007).

[47] Firstenberg, O., Adams, C. S., & Hofferberth, S. Nonlinear quantum optics mediated by Rydberg interactions. *J. Phys. B* **49**, 152003 (2016).

[48] Huang, K.-Y. *et al.* Quantum squeezing of vector slow-light solitons in a coherent atomic system. *Chaos, Solitons and Fractals* **163**, 112557 (2022).

[49] Hang, C., Li, W., & Huang, G. Nonlinear light diffraction by electromagnetically induced gratings with PT symmetry in a Rydberg atomic gas. *Phys. Rev. A* **100**, 043807 (2019).

[50] Xu, S.-L., Zhou, Q., Zhao, D., Belić, M. R., & Zhao, Y. Spatiotemporal solitons in cold Rydberg atomic gases with Bessel optical lattices. Appl. Math. Lett. 106, 106230 (2020).

[51] Gorshkov, A. V., Otterbach, J., Fleischhauer, M., Pohl, T., & Lukin, M. D. Photon-Photon Interactions via Rydberg Blockade. *Phys. Rev. Lett.* **107**, 133602 (2011).

[52] Guo, Y.-W., Xu, S.-L., He, J.-R., Deng, P., Belić, M. R., & Zhao, Y. Transient optical response of cold Rydberg atoms with electromagnetically induced transparency. *Phys. Rev. A* **101**, 023806 (2020).

[53] Li, B.-B. *et al.* Two-Dimensional Gap Solitons in Parity-Time Symmetry Moiré Optical Lattices with Rydberg-Rydberg Interaction. *Chin. Phys. Lett.* **40**, 044201 (2023).

[54] Q.-Y. Liao *et al.* Two-dimensional spatial optical solitons in Rydberg cold atomic system under the action of optical lattice. *Acta Physica Sinica* **72**, 104202 (2023).

[55] Gallagher, T. F. Rydberg Atoms (Cambridge University press, England, 2008).

[56] Maxwell, D. *et al.* Storage and control of optical photons using Rydberg polaritons. *Phys. Rev. Lett.* **110**, 103001 (2013).

[57] Yang, L., He, B., Wu, J.-H., Zhang, Z., & Xiao, M. Interacting photon pulses in a Rydberg medium. *Optica* **3**, 1095-1103 (2016).

[58] Baur, S., Tiarks, D., Rempe, G., & Dürr, S. Single-photon switch based on

Rydberg blockade. *Phys. Rev. Lett.* **112**, 073901 (2014).

[59] Gorniaczyk, H. *et al.* Enhancement of Rydberg-mediated single-photon nonlinearities by electrically tuned Förster resonances. *Nat. Commun.* **7**, 12480 (2016).

[60] Li, L., & Kuzmich, A. Quantum memory with strong and controllable Rydberg-level interactions. *Nat. Commun.* **7**, 13618 (2016).

[61] Bychkov, Yu. A. & Rashba, E. I. Oscillatory effects and the magnetic susceptibility of carriers in inversion layers. *J. Phys. C: Solid State Phys.* **17**, 6039-6045 (1984).

[62] Wang, Y.-J., Wen, L., Chen, G.-P., Zhang, S.-G. & Zhang, X.-F. Formation, stability, and dynamics of vector bright solitons in a trapless Bose–Einstein condensate with spin–orbit coupling. *New J. Phys.* **22,** 033006 (2020).

[63] Henkel, N., Nath, R. & Pohl, T. Three-Dimensional Roton Excitations and Supersolid Formation in Rydberg-Excited Bose-Einstein Condensates. *Phys. Rev. Lett.* **104**, 195302 (2010).

[64] Chen, Z., Li, Y., Malomed, B. A., & Salasnich, L. Spontaneous symmetry breaking of fundamental states, vortices, and dipoles in two- and one-dimensional linearly coupled traps with cubic self-attraction. *Phys. Rev. A* **96**, 033621 (2017).

[65] Chen, X.-Y. *et al.* Magic tilt angle for stabilizing two-dimensional solitons by dipole-dipole interactions. *Phys. Rev. A* **96**, 043631 (2017).

[66] Bai, Z., Li, W., & Huang, G. Stable single light bullets and vortices and their active control in cold Rydberg gases. *Optica* **6**, 309-317(2019).

[67] Yang, J. & Lakoba, T. I. Universally-convergent squared-operator iteration methods for solitary waves in general nonlinear wave equations. *Stud. Appl. Math.* **118**, 153-197 (2007).

[68] Zhang, X.-F., Kato, M., Han, W., Zhang, S.-G., & Saito, H. Spin-orbit-coupled Bose-Einstein condensates held under a toroidal trap. *Phys. Rev. A* **95**, 033620 (2017).

[69] Sakaguchi, H. & Malomed, B. A. Solitons in combined linear and nonlinear lattice potentials. *Phys. Rev. A* **81**, 013624 (2010).